\documentclass[twocolumn,preprintnumbers,superscriptaddress,amsmath,amssymb]{revtex4}
\usepackage{color}
\usepackage{graphics}
\usepackage{dcolumn}
\usepackage{bm}

\renewcommand\textfraction 0
\renewcommand\topfraction 1
\renewcommand\bottomfraction 1
\begin{document} 

\preprint{Thin Solid Films, 453-454 (2004) 557-561}

\title{CLUSTER EMISSION UNDER FEMTOSECOND LASER ABLATION OF SILICON}

\author{Alexander V. \surname{Bulgakov}}
\email{bulgakov@itp.nsc.ru  (A.V. Bulgakov)}
\affiliation{Institute of Thermophysics SB RAS, Prospect Lavrentyev 1, 630090 
Novosibirsk, Russia}

\author{Igor \surname{Ozerov}}
\affiliation{
GPEC, UMR 6631 CNRS, Facult\'{e} des Sciences de Luminy, Case 901, 
13288 Marseille Cedex 9, France.}

\author{Wladimir \surname{Marine}}
\altaffiliation{Corresponding author:\\
Phone: +33 4 91 82 91 73\\ Fax: +33 4 91 82 91 76\\
Electronic address: marine@crmcn.univ.mrs.fr  (W. Marine)}
\affiliation{
GPEC, UMR 6631 CNRS, Facult\'{e} des Sciences de Luminy, Case 901, 
13288 Marseille Cedex 9, France.}

\begin{abstract}

Rich populations of clusters have been observed after femtosecond laser 
ablation of bulk silicon in vacuum. Size and velocity distributions of the 
clusters as well as their charge states have been analyzed by reflectron 
time-of-flight mass spectrometry. An efficient emission of both neutral 
silicon clusters Si$_{n}$ (up to n = 6) and their cations Si$_{n}^{ + }$ 
(up to n = 10) has been observed. The clusters are formed even at very low 
laser fluences, below the ablation threshold, and their relative yield increases 
with fluence. We show the dependencies of the cluster yield as well as the 
expansion dynamics on both laser wavelength and laser fluence. The 
mechanisms of the cluster formation are discussed.


\end{abstract}
\maketitle

The recent development of femtosecond lasers expands the possible 
applications of laser ablation. However, the fundamental mechanisms of 
light-material interactions are still poorly understood. A study of the 
plume composition and its expansion helps in understanding of the 
fundamental processes involved into the interaction and in the development 
of new laser applications. Recently, surface modifications under femtosecond 
laser irradiation of bulk silicon have been analyzed for laser pulse 
durations from 5 to 400 fs \cite{1}. The pump-and-probe studies show the moment 
of the plasma initiation and allow one to follow the plume expansion at the 
very early stages close to the target \cite{2}. The first observation of the 
silicon plume expansion at larger time scale after the femtosecond ablation 
by time-of-flight methods was reported in \cite{3}, where the authors also 
mentioned the observation of small nanoclusters. However, very few 
information is available on the formation of silicon nanoclusters after the 
ablation by short laser pulses. The dimer desorption under resonant 
nanosecond excitation has been recently reported in the case of (2x1) 
reconstructed (100) silicon surface \cite{4}. The use of high energy photons of 
6.4 eV (ArF laser) induces very strong desorption and ablation of atomic 
species and small clusters from both the bulk material and the 
nanostructured Si surfaces \cite{5}. However, there is no systematical 
information neither on composition nor on the dynamical properties of the 
plumes induced by femtosecond irradiation. In this work, we present evidence 
for the formation of clusters with sizes up to 10 atoms under ultrashort 
laser ablation of bulk silicon and analyze the dynamics of their expansion.

\section{Experimental}

The experiments were performed with Si [100] surface under ultrahigh vacuum 
conditions ($\sim $ 10$^{-10}$ mbar). The monocrystalline Si target was 
irradiated at an angle of incidence of 45\r{ } using a Ti:sapphire laser 
system (Mai-Tai coupled with TSA amplifier, Spectra Physics, 80 fs pulse 
duration, 10 Hz repetition rate, up to 30 mJ energy per pulse) operating at 
wavelengths of 800, 400, and 266 nm. A part of the laser beam was selected 
by an aperture to provide a nearly uniform intensity distribution over the 
irradiated spot. The target was rotated/translated during measurements to 
avoid considerable cratering. For each wavelength, series of craters at 
several laser energies were produced on the stationary sample. By measuring 
the crater areas, the absolute calibration of laser fluence was performed. 
The laser fluence on the target was varied in the ranges 80$ -$800, 20$ - 
$200, 5$ -$50 mJ/cm$^{2}$ at wavelengths of 800, 400, and 266 nm, 
respectively. 

The abundance distributions of neutral and cationic particles in the 
laser-induced plume were analyzed by a reflectron time-of-flight mass 
spectrometer (TOF MS). When ions were studied, the plume was allowed to 
expand under field-free conditions. While the neutral particles of the plume 
were analyzed, the ionized species were rejected using a simple plasma 
suppressor \cite{6}. The latter consisted of a pair of deflection plates placed 
along the plume axis in front of the ion source of the MS where neutral 
particles were ionized by impact of 110 eV electrons. At a distance of 11 cm 
from the target, the ions, either ablated or post-ionized, were sampled 
parallel to the plume expansion axis by a 500 V repeller pulse after a delay 
time $t_{d}$, following the laser pulse. All mass spectra were averaged over 
300 laser shots.

\section{Results and discussion}

\begin{figure}[!]
 \resizebox{8.5cm}{!}{\includegraphics{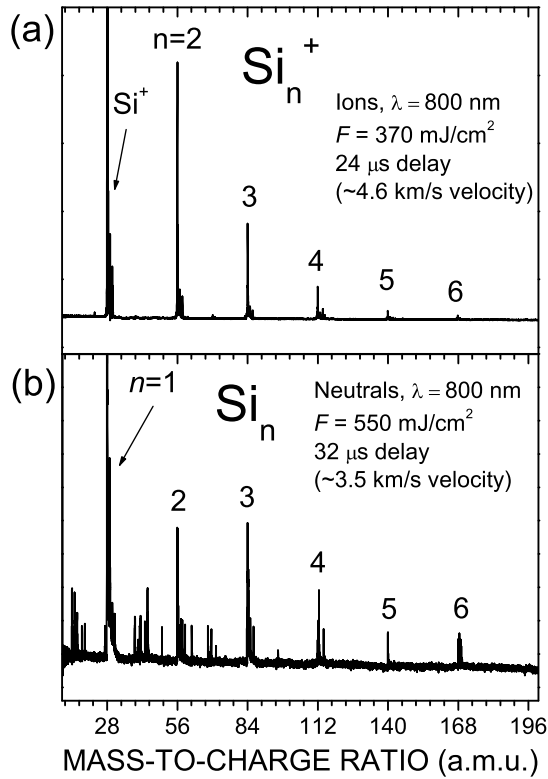}} \caption{\label{fig1}
Mass spectra of cationic (a) and neutral (b) species. The 
irradiation conditions are indicated in the figure.}
\end{figure}

Efficient emission of both neutral silicon clusters Si$_{n}$ (up to $n$ = 6) 
and their cations Si$_{n}^{ + }$ (up to $n$ = 10) have been observed for all 
investigated wavelengths. Si atoms and Si$^{ + }$ ions were the most 
abundant particles at all laser fluences. Though, at certain irradiation 
regimes, the cluster fraction exceeded 10{\%} of that of Si atoms and ions 
in the ablation plume. Fig. \ref{fig1} shows typical mass spectra of cationic and 
neutral plume species, respectively. The spectra were obtained under the 
irradiation conditions corresponding to the maximum relative yield of the 
clusters with respect to monatomic particles. The abundance distributions 
are smooth under these conditions with peak intensities monotonously 
decreasing with cluster size. As it will be discussed below, the velocity 
distributions for ionized plume particles are strongly dependent on particle 
mass. The ion mass spectra are ,therefore, different for particles expanding 
in the faster and the slower parts of the plume. The mass spectrum shown in 
Fig. \ref{fig1}a was obtained at the time delay corresponding to the maximum yield of 
Si$_{2}^{ + }$ dimer.

\begin{figure}[!]
 \resizebox{8.5cm}{!}{\includegraphics{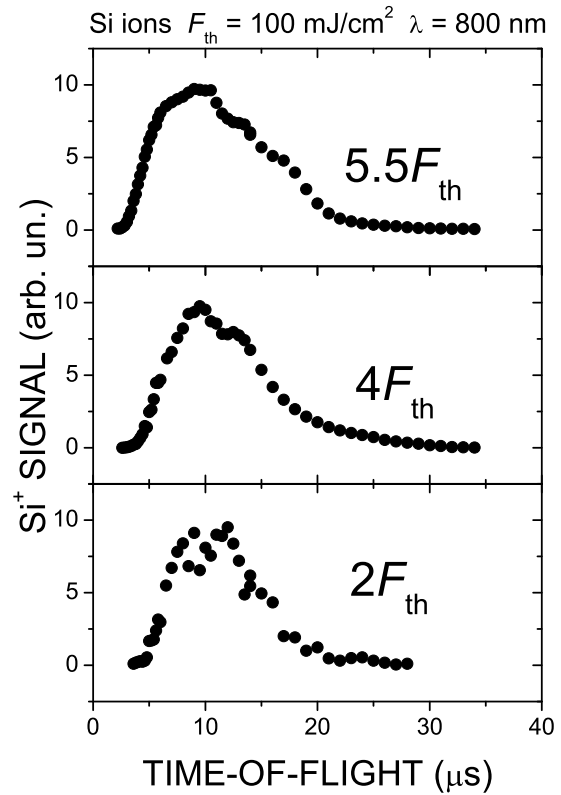}} \caption{\label{fig2}
Typical time-of-flight (TOF) distributions of Si$^{ + }$ ions for 
different fluences. Laser wavelength was 800 nm. $F_{th }$= 100 mJ/cm$^{2}$ 
is the threshold fluence corresponding to the ion appearance in the plume.}
\end{figure}

A variation in the time delay $t_{d}$ allowed the analysis of particle 
velocity distributions and the characterization of the temporal evolution of 
the plume composition. Figure \ref{fig2} shows typical TOF distributions of Si$^{ + 
}$ ions for different fluences. We have found that the distributions are 
weakly affected by both laser wavelength and fluence. In all cases, they 
reach the maximum at $\sim $9 $\mu $s time delay that corresponds to the ion 
velocity of around 12 km/s, or to a kinetic energy of $\sim $20 eV. An 
increase in laser fluence results only in a broadening of the distributions 
with a negligible shift towards higher velocities. At the same time, the 
total number of Si$^{ + }$ ions increases strongly with fluence. The ion 
yield at a given fluence can be evaluated by integrating the TOF 
distributions in time. To obtain the values proportional to the total number 
of the plume particles at various fluences, this procedure should be 
corrected for angular distributions of expanding particles as a function of 
their velocity (usually fast plume species are more forward directed than 
slow ones \cite{7,8}). The angular distributions were not measured in this work, 
but, since the ion velocity distributions are similar for different 
conditions (Fig.\ref{fig2}), such a correction is not necessary for Si$^{ + }$ ions. 

\begin{figure}[!]
 \resizebox{8.5cm}{!}{\includegraphics{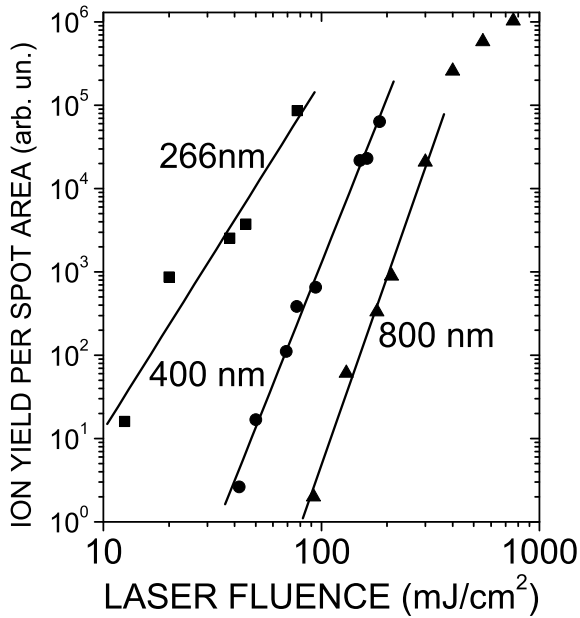}} \caption{\label{fig3}
The ion yield as a function of laser fluence for 800, 400 and 266 nm 
wavelengths of irradiation.}
\end{figure}

Figure \ref{fig3} shows the yields of plume ions as a function of laser fluence for 
three laser wavelengths. These dependencies show a steep gradient that is 
nearly identical for 800 and 400-nm wavelengths. For 266-nm, it is slightly 
weaker. We note that the total number of detected particles varies by around 
6 orders of magnitude in the studied fluence ranges. The threshold fluences, 
$F_{th}$, for ion appearance in the plume, as deduced from MS measurements 
are around 100, 40, and 10 mJ/cm$^{2}$ for 800, 400, and 266 nm, 
respectively. It should be noted that our ion detection system has a high 
sensitivity to detect a single plume ion. As a result, the obtained values 
correspond to the \textit{real} thresholds for ion appearance but not to the ion 
\textit{detection} threshold. 

Cluster ions Si$_{n}^{ + }$ were observed throughout the laser fluence 
range studied, even at very low fluences close to the threshold for Si ions 
(at least for dimers). The clusters are found to be considerably slower than 
atomic ions. At near-threshold fluences, the TOF distributions for Si$_{2}^{ + }$ dimers are maximized 
at $t_{d}\cong $18 $\mu $s for all wavelengths. The most probable 
velocity of the dimers is, therefore, approximately two times lower than that 
of Si$^{ + }$. This means that these plume particles have nearly the same 
momentum under these conditions. However, the Si$_{3}^{ + }$ trimers are 
faster than it would be expected for constant momentum (the TOF 
distributions are maximized at around 21 $\mu $s).

By integrating the cluster TOF distributions in time, the total number of 
clusters in the plume as a function of fluence was also evaluated. For low 
fluences, the total number of the detected dimers increases with fluence and 
reaches maximum at $F$/$F_{th}\cong $ 2 for 800 nm and at $F$/$F_{th} \cong $ 
4 for 400 nm. The maximum values of the population ratios Si$_{2}^{ + 
}$/Si$^{ + }$ are around 0.18 and 0.22 for 800 and 400 nm respectively. 
Further increase in fluence results in an abrupt fall of the relative 
concentration of the dimers. This decrease of the Si$_{2}^{ + }$ abundance 
in the plume is due to both overall decrease of the relative cluster yield 
at high fluences and to a conversion of Si$_{2}^{ + }$ into larger 
cluster. The performed TOF integration shows the fast increase of the 
Si$_{4}^{ + }$ and Si$_{6}^{ + }$ cluster abundance with fluence at $F >$ 400 mJ/cm$^{2}$.

\begin{figure}[!]
 \resizebox{8.5cm}{!}{\includegraphics{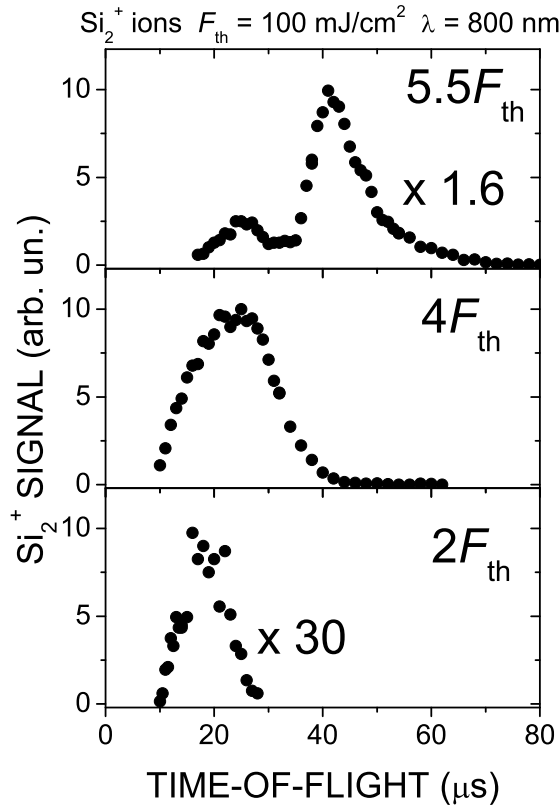}} \caption{\label{fig4}
TOF distributions of Si$_{2 }$for irradiation with different 
fluences at laser wavelength of 800 nm.}
\end{figure}

In contrast to Si$^{ + }$ ions, the TOF distributions of clusters strongly 
depend on the laser fluence. Figure \ref{fig4} shows the evolution of the TOF 
distribution for the Si$_{2}^{ + }$ dimer at different fluences. These 
results were obtained for excitation at 800 nm. At very low fluences, below 
$\sim $ 200 mJ/cm$^{2}$ ($ \cong $ 2$F_{th})$, the distributions are rather 
narrow, single-peaked, and maximized at $\sim $18 $\mu $s. For laser 
fluences above 2$F_{th}$ , the behavior of Si$_{2}^{ + }$ changes 
dramatically. In the fluence range between 2$F_{th}$ and 4$F_{th}$, the TOF 
distributions are still single-peaked but their maxima are 
significally shifted towards higher time delays (lower velocities). At $F = $4$F_{th}$, 
the maximum is at $\sim $25 $\mu $s (the corresponding most probable 
velocity is $\sim $ 4.4 km/s). The fast cluster ions are still present in 
this ablation regime, as can be seen in Fig. \ref{fig4}, but their distribution is 
masked by slower ions. With further increase in laser fluence the second 
slower population of Si$_{2}^{ + }$ ions with the most probable velocity 
of around 2.7 km/s appears in the plume. At fluences above $\sim $5.5$ F_{th}$, 
this slow population becomes dominant. The first faster peak in the 
distribution (corresponding to the cluster ions detected in the (2$-$4)$ F_{th}$ 
fluence range) is still present, well separated from the second peak, and is 
still maximized at $\sim $25 $\mu $s. However, its intensity is much lower 
then at low fluences. The fast cluster ions (those observed at $F <$ 
2$F_{th})$ are not detected anymore. 

The fluence dependency of the TOF distributions for larger Si$_{n}^{ + }$ 
clusters ($n$ = 3$-$6) under under 800-nm excitation is found 
to be qualitatively similar to that for Si$_{2}^{ + }$. At a threshold 
fluence of around 450 mJ/cm$^{2}$, the distributions are transformed from 
the single-peaked one to the double-peaked with the second (slow) cluster 
population becoming rapidly dominant with further increase in fluence. The 
most probable velocity of the second population slightly decreases with 
cluster size from 2.5 km/s for Si$_{2}^{ + }$ to 2 km/s for Si$_{6}^{ + 
}$. The abundance distribution for this slow cluster population is 
remarkably different from that for fast clusters (shown in Fig. \ref{fig1}a). The 
Si$_{4}$ and Si$_{6}$ cluster ions are more abundant than their odd-numbered 
neighbors. For larger time delays, the Si$_{6}$ cluster becomes dominant 
particle in the plume.

Based on the obtained results, we can address the fundamental questions on 
the mechanism of cluster formation. Are the observed clusters formed 
primarily by their direct ejection from the target or by gas-phase 
aggregation of ablated atoms? Which process is responsible for such a 
transformation of the cluster velocity distribution as shown in Fig. \ref{fig4}? The 
set of the obtained TOF distributions and fluence dependencies provides a 
clear evidence for cluster formation mechanisms at different irradiation 
regimes. At very low laser fluence, from ion appearance threshold $F_{th}$ up 
to $\sim $2$ F_{th}$, Si$^{ + }$ and Si$_{2}^{ + }$ ions are produced by an 
impulsive Coulomb explosion (CE) from a charged surface. We argue this by 
the following:

(a) The observed momentum scaling for the particles and a very weak 
dependence of their expansion velocities on laser fluence and wavelength in 
the low fluence regimes indicates directly the CE mechanism \cite{7}. The 
repulsive electric field induced in the target by the fs-laser pulse lasts 
for a short period of time ($\sim $1 ps). Both Si$^{ + }$ and Si$_{2}^{ + 
}$ spend roughly the same time in the action range of this field. The time 
of electrostatic interaction leading to the ion removal is thus many orders 
of magnitude shorter than the ion flight time to the mass spectrometer that 
results in equal momenta for the ejected Si$^{ + }$ and Si$_{2}^{ + }$ 
ions as seen experimentally (Fig.\ref{fig4}). We suggest that the electric field 
under these conditions is strong enough to repulse positive ions from the 
surface but not so strong to break the bonds of ejected clusters. Note that 
thermally desorbed particles are characterized by equal kinetic energy 
rather than momentum \cite{7}.

(b) An additional support for the CE is the high velocity of Si$_{2}^{ + 
}$ ions at low fluences. For fluences beyond $\sim $2$F_{th}$, the most 
probable velocity decreases that indicate the appearance of an additional 
ablation channel. For thermally desorbed particles one would expect the 
opposite behavior with fluence.

Contrarily, the slower Si$_{n}^{ + }$ clusters (which overshadow the 
Coulomb explosion ions in the 2$F_{th} -$4$F_{th}$ fluence range and form the 
fast cluster population at higher fluences) can be interpreted as "plasma 
ions", i.e., ions formed in the gas-phase ionized vapor plume. Also, we 
should take in consideration the direct thermal evaporation of small 
clusters from the surface (desorption of Si$_{n}$ clusters up to $n$ = 6 was 
observed from thermally heated Si \cite{9}). The velocities of these "plasma 
ions" decrease only slightly with cluster size and scale by a law 
intermediate between constant kinetic energy and constant velocity.

\section{Conclusions}

In conclusion, we present the first analysis of expansion dynamics of 
monoatomic species and small clusters after femtosecond laser ablation of 
clean (100) Si surface. While our results unambiguously show the domination 
of nonthermal mechanism of nanocluster emission at low laser fluence, more 
studies is needed to be certain about the origin of the ions under all the 
irradiation conditions.

\end{document}